\documentclass{article}

 \usepackage[final]{neurips_mlsb_2025}
 \usepackage[inkscapelatex=false]{svg}
 \usepackage{multirow}

\usepackage[utf8]{inputenc} 
\usepackage[T1]{fontenc}    
\usepackage{hyperref}       
\usepackage{url}            
\usepackage{booktabs}       
\usepackage{amsfonts}       
\usepackage{nicefrac}       
\usepackage{microtype}      
\usepackage{xcolor}         
\usepackage{graphicx}
\usepackage{float}
\usepackage{amsmath}

\title{Improving scoring functions for protein-protein docking with LambdaLoss}

\author{
  Richard Zhu \\
  Department of Statistics \\
  Harvard University \\
  \texttt{rzhu@college.harvard.edu}
  \And
  Darren Xu, Lee-Shin Chu \& Jeffrey J. Gray \\
  Department of Chemical and Biomolecular Engineering \\
  Johns Hopkins University \\
  \texttt{\{dxu39,lchu11\}@jh.edu, jgray@jhu.edu}
}

\begin{document}

\maketitle

\begin{abstract}
  Modeling protein-protein interactions requires accurate scoring functions that can rank potential poses (conformations) of a protein-protein complex to differentiate near-native poses from incorrect ones. Here, we propose a general framework for improving protein-protein pose ranking and other biomolecular interaction models using the LambdaLoss loss function from the Learning-to-Rank field. We test this framework by fine-tuning the energy prediction head of DFMDock with the LambdaLoss on an augmented dataset of 2.9M decoy poses derived from the DIPS dataset. On targets from the CAPRI score set benchmark, our fine-tuned ranking model LambdaDockScore is better at identifying correct poses in its top-1 and top-5 predictions compared to EuDockScore, a state-of-the-art method. LambdaDockScore also improves upon baseline DFMDock ranking performance for scoring antibody-antigen complexes and protein-protein complexes with very large or small binding interfaces.
\end{abstract}

\section{Introduction}
Protein-protein interactions are ubiquitous in biology, underlying many fundamental processes such as the recognition of viral antigens by human antibodies, the function of molecular machines like ATP synthase, and the conformational changes that drive signal transduction pathways. Recent advances in biomolecular modeling, such as AlphaFold3, have improved our capacity to model protein-protein interactions [1,13]. However, certain interactions remain challenging to model, such as predicting the interactions between antibodies and antigens [5,12,20].

Modeling protein-protein interactions is often decomposed into two subprocesses: 1.\ sampling plausible poses (conformations) of the protein-protein complex and then 2.\ ranking the generated poses to find the correct one, usually based on scoring functions that describe the biophysical quality of the poses. Thus, the development of accurate scoring functions is critical for success in modeling protein-protein interactions. Recently, deep-learning-based scoring functions have been shown to outperform classical methods and set a new bar in identifying near-native poses [14]. However, they can still struggle to correctly rank the poses of difficult protein-protein complexes [14].

In this study, we propose a novel framework for improving deep-learning-based scoring functions for biomolecular interactions. While we focus on protein-protein interactions here, this technique could be directly applied to other interactions as well. We employ a careful data augmentation process to generate a large dataset of protein-protein poses that spans a wide spectrum of biophysical quality, and then we fine-tune a base ranking model on our augmented dataset with the LambdaLoss. LambdaLoss is a machine learning loss function used in ranking query results in information retrieval systems like search engines [3,19].

We test our framework on the energy prediction head of DFMDock, a unified sampling and ranking model for protein-protein docking whose training objective encourages learning the underlying energy landscape of protein-protein interactions [4]. We fine-tune this base model using the LambdaLoss and our augmented dataset to create the LambdaDockScore ranking model. On a set of 69 targets in the CAPRI score set [8], LambdaDockScore outperforms baseline DFMDock and a state-of-the-art ranking model EuDockScore in identifying the correct poses among the top-1 and top-5 ranked poses. We analyze LambdaDockScore's performance on subtypes of protein-protein complexes using another test set generated by running DFMDock inference on the Docking Benchmark 5.5 (DB5.5) [18]. In this test set, LambdaDockScore demonstrates the largest improvements relative to the baseline DFMDock ranking performance for two subtypes: antibody-antigen complexes and protein-protein complexes with very large or very small binding interfaces.


\section{Related work}

\subsection{Modeling protein-protein interactions with deep learning}
\label{sec:supp_ppi}
Diffusion-based deep learning models have shown great promise in protein-protein docking, which aims to predict the structures of bound protein complexes from the structures of unbound monomers. For example, DiffDock-PP is a diffusion model that performs rigid-body docking: it performs only translations and rotations (with no protein backbone or side chain movements) to predict the bound complex. DiffDock-PP set the baseline accuracy for deep learning methods on the Database of Interacting Protein Structure (DIPS) dataset [6]. Building on this, DFMDock proposed the first diffusion-based docking model to unify sampling and ranking in one framework: it simultaneously learns translation and rotation forces to sample poses of the bound complex and an energy function to score these poses. DFMDock demonstrates better generalization than DiffDock-PP with superior performance on standard benchmarks (\textit{i.e.} DB5.5) and a learned energy function that exhibits realistic binding funnels [4]. Most diffusion-based docking methods are paired with separate confidence models for ranking [6]; DFMDock’s innovation lies in using its own energy prediction head for ranking, eliminating the need for an external scoring model [4].

In parallel, large ``co-folding'' models have emerged that jointly predict the complex structure of two interacting proteins from their sequences. AlphaFold3 uses a diffusion module for structure prediction and can model interactions between proteins, nucleic acids, small molecules, and ions [1]. Boltz-2 is another all-atom co-folding model that builds on the architecture of AlphaFold to incorporate additional functionalities like method, contact, and pocket conditioning [13]. These models outperform diffusion-based docking models, setting the state-of-the-art in deep learning for protein-protein complex prediction.  However, these co-folding approaches rely heavily on evolutionary information extracted from multiple sequence alignments (MSAs) of related proteins. When homologous proteins are scarce or the evolutionary process is unique---as in many antibody--antigen complexes---their performance degrades [5,12,20]. Like diffusion models, co-folding models generate candidate poses and then use internal confidence or scoring metrics to rank them. For instance, AlphaFold3 and Boltz-2 produce confidence metrics like pLDDT and PAE [1,13].

Thus, virtually all current methods sample protein-protein complex poses and use a separate confidence model to predict the biophysical quality of poses for ranking. In the case of DFMDock, the learned energy from its energy prediction head is used for ranking instead of a separate model.

\subsection{Scoring functions}
\label{sec:supp_score}
Scoring functions take a proposed protein-protein complex and predict its biophysical quality (\textit{e.g.} how close it is to the native complex). In docking pipelines, this is typically the second step after sampling poses: the scoring function should assign high scores to near-native poses and low scores to incorrect ones.

In recent years, deep learning approaches have outperformed classical scoring functions on multiple protein-protein docking benchmarks [14]. Current state-of-the-art deep learning scoring models include PIsToN (Protein Interfaces with Transformer Network), which crops protein-protein interface patches into images and processes them with a vision transformer [15]; dMaSIF, which represents the protein surface as a point cloud and applies geodesic convolutions to compute an interface binding score [16]; and EuDockScore, which uses an SE(3)-equivariant graph neural network to score protein--protein interfaces [10]. These methods learn to use geometric, chemical and/or energy features via neural networks, allowing them to set a new standard in scoring functions for protein-protein docking.

\subsection{Learning-to-Rank methods}
\label{sec:supp_ltr}
LambdaLoss originates from the Learning-to-Rank field, where models are trained to optimize ranking metrics (\textit{e.g.} normalized discounted cumulative gain (NDCG)) rather than simple regression or classification losses [9]. The LambdaLoss loss function optimizes an upper bound on the NDCG metric [19]. Given a list of entities that must be ranked by their relevance to the query, the LambdaLoss analyzes pairs of entities, where the pairwise terms depend on both the ground-truth ranks and the difference in true relevance scores [19]. This formulation bridges rank-based and score-based ranking objectives and demonstrates state-of-the-art performance on Learning-to-Rank benchmarks [3,19].

\section{Method}

\subsection{Applying LambdaLoss for biomolecular ranking}
To adapt the LambdaLoss ranking framework to scoring biomolecular interactions, we modify the original LambdaLoss objective as follows. For each biomolecular complex (protein-protein complex, protein-nucleic acid complex, \textit{etc.}) in the training set, we have a set of sampled poses $P = \{p_1, ...,p_N\}$. For pose $p_k \in P$, its ground truth quality score is denoted as $q_k$, its predicted quality score from the ranking model is denoted as $s_k$, and its rank in the pose list sorted by ground truth quality score is denoted as $\mathrm{rank}(k)$. Then, we can define the biomolecular interaction LambdaLoss as the sum of pairwise loss terms between each pair of poses $p_i, p_j \in P$ for which $q_i > q_j$, as shown in Equation \ref{eq:lambdaloss}. This captures the rank- and score-weighted difference between the ground truth and predicted lists of pose rankings.

\begin{equation}
\label{eq:lambdaloss}
\mathcal{L} =
\mathbb{E} \left[
  \sum_{q_i > q_j}
    \underbrace{\Delta_{q}\, \Delta_{\mathrm{rank}}}_{\text{fixed weights}} \,
    \log \left(1 + e^{-(\overbrace{s_i - s_j}^{\text{learned scores}})}\right)
\right]
\end{equation}

Here, $\mathbb{E}$ denotes the expectation taken over all biomolecular complexes in the training set, while the inner summation aggregates the loss over all valid pose pairs $(p_i, p_j)$ for a given complex where $q_i > q_j$. Within the summation, $\Delta_{q}$ and $\Delta_{\mathrm{rank}}$ are fixed weights calculated for each pair, whereas $s_i - s_j$ is a difference in the learned scores that the model actively optimizes.

The weighting terms in Equation \ref{eq:lambdaloss} are defined as
\begin{equation}\label{eq:deltas}
\Delta_{q} = \left|2^{q_i}-2^{q_j}\right|,\qquad
\Delta_{\mathrm{rank}} = \left|\frac{1}{\log(1+\mathrm{rank}(i))}-\frac{1}{\log(1+\mathrm{rank}(j))}\right|
\end{equation}

Together, these equations define the biomolecular interaction LambdaLoss training objective. 

The rationale behind using this specific loss function is to prioritize the relative ordering of the most relevant poses. Specifically, the $\Delta_q$ weight exponentially penalizes incorrect ordering between poses with large differences in ground-truth quality, forcing the model to distinguish near-native complexes from poor ones. Meanwhile, the $\Delta_{\mathrm{rank}}$ weight heavily penalizes ranking errors at the top of the pose list—which contain the absolute best poses—because the reciprocal weighting places greater emphasis on higher-ranked poses.

\subsection{Fine-tuning DFMDock} We fine-tune DFMDock's protein-protein scoring ability using the biomolecular interaction LambdaLoss in Equation \ref{eq:lambdaloss}. For this specific application, the ground truth quality metric $q$ is the protein-protein complex pose's DockQ score, a measure of distance from the native complex structure [2]. The predicted quality score $s$ is $-E$, where $E$ is the energy output by DFMDock's energy prediction head. Thus, a lower energy corresponds to a higher predicted score and a prediction of higher pose quality. The DFMDock base model was chosen because its original training objective encourages it to learn the energy landscape of protein-protein docking by optimizing its energy head to predict a lower energy for the ground truth pose (the experimentally-determined native state) compared to the predicted pose. One limitation of this objective is that only two poses are compared at a time, and one of them is always the ground truth. By fine-tuning the model with our new loss function, we further refine the landscape by providing 270 non-native-state decoys per complex in the training dataset and applying the LambdaLoss to encourage the model to rank these decoys (and the ground truth pose) correctly by their DockQ scores. We refer to DFMDock before fine-tuning as the ``baseline'' model (which includes both a pose sampler and energy prediction head), while the fine-tuned energy prediction head is referred to as our ``LambdaDockScore'' model.


The 270 decoys per complex in the training set were generated as follows: 220 complexes were generated by perturbing the ground truth protein-protein complex with varying levels of noise, while 50 complexes were generated as DFMDock predictions for the complex structure (Supplementary Table \ref{tab:decoy-distribution}). The ground truth pose was also included for a total of 271 poses per complex. For each protein complex in the training set, we used this strategy to generate a set of poses spanning the $[0,1]$ range of DockQ quality scores. Supplementary Figure \ref{fig:fig_datadist} shows that most poses with DockQ $<$ 0.23 are generated from DFMDock sampling, while most poses with DockQ $\geq$ 0.23 are generated from perturbations of the ground truth. This wide range in pose quality is important for the success of the data-augmentation-based fine-tuning strategy for training LambdaDockScore, as it allows the model to rank a wide range of pose qualities encountered at inference time. The final fine-tuning dataset contains 2,906,199 decoy poses for 10,724 protein-protein complexes from the DIPS-hetero training dataset (a heterodimer subset of the DIPS dataset). Validation was performed on the DIPS-hetero validation dataset.

After generating the fine-tuning dataset, we performed full fine-tuning of the whole DFMDock model using its energy predictions and the LambdaLoss loss function shown in Equation \ref{eq:lambdaloss}. The full fine-tuning loss was a weighted combination of the original DFMDock training loss [4] and the LambdaLoss presented in Equation \ref{eq:lambdaloss}:
\begin{equation}\label{eq:loss}
\mathcal{L}_{\mathrm{total}} = 0.5\times \mathcal{L}_{\mathrm{original}} + 5.0\times \mathcal{L}_{\mathrm{lambdaloss}}
\end{equation}
The fine-tuning was conducted for 36 epochs with a learning rate of $1 \times 10^{-4}$. For each protein complex in each training epoch, ten decoys were sampled for ranking that spanned the entire range of DockQ scores. This was achieved by dividing the DockQ $[0,1]$ range into ten buckets of width $0.1$, assigning each pose a sampling probability of $\frac{1}{\text{\# poses in the bucket}}$, and sampling ten poses from the resulting distribution. On average, each bucket had one pose sampled, allowing the model to learn a wide range of DockQ scores.

\subsection{Model evaluation procedures} 

For evaluation, LambdaDockScore was used either as a standalone scoring function for existing decoy poses (CAPRI score set) or paired with the baseline DFMDock sampling model for test datasets without existing decoy poses (DB5.5). In both cases, the top-1 and top-5 success rate metrics were calculated, which entails identifying the top-$k$ decoy poses with the lowest predicted energy from the LambdaDockScore output and calculating their ground truth DockQ values. The top-$k$ success rate is then the percent of protein-protein complexes in the test set where at least one of the top-$k$ decoy poses ranked by LambdaDockScore surpasses the DockQ thresholds for acceptable- ($\text{DockQ} >0.23$), medium- ($\text{DockQ} >0.49$), or high-quality ($\text{DockQ} >0.80$) poses, as defined by the Critical Assessment of PRedicted Interactions (CAPRI) initiative [2,7].

\section{Results}

\subsection{LambdaDockScore outperforms a state-of-the-art protein-protein interface scoring function on the CAPRI score set}

\begin{figure}[h]
  \centering
  \includegraphics[width=0.85\textwidth]{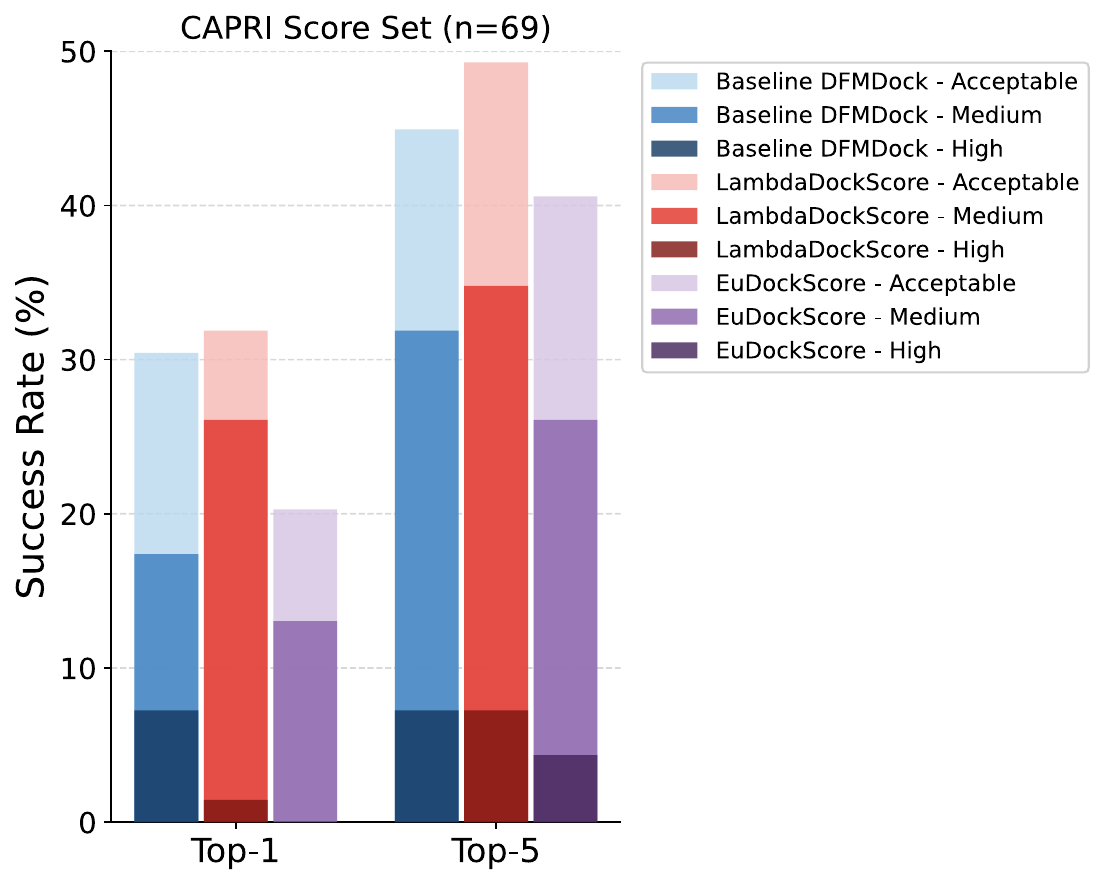} 
  \caption{Benchmarking results on 69 protein-protein complexes from the CAPRI Score Set (filtered for <30\% seq. identity to training data). Each complex averaged ${\sim}$1.1k decoys. The top-1 and top-5 ranked decoys from each model were analyzed. Exact success rates for all models are provided in Supplementary Table \ref{tab:capri_bar_heights}.}
  \label{fig:fig_capri}
\end{figure}

In Figure \ref{fig:fig_capri}, we compare the scoring ability of LambdaDockScore (red) with the pose-scoring energy prediction head of the baseline DFMDock model (blue) and a state-of-the-art protein-protein ranking model EuDockScore (purple), which uses an SE(3)-equivariant graph neural network for protein--protein interface scoring. Since DFMDock and LambdaDockScore were trained on the DIPS-hetero subset of the DIPS dataset [11,17], while EuDockScore was trained on DB5.5 [18], the CAPRI score set was chosen as a challenging, independent evaluation set for all three models [8]. To address data leakage, we filtered complexes from the CAPRI score set (v2022) against the DIPS-hetero training and validation sets using a 30\% sequence identity cutoff. After filtering and excluding three-chain complexes, we obtained a final test set of 69 protein-protein complexes with an average of 1,103 decoy poses each. The three models were tasked with ranking all decoy poses for each protein-protein complex, and the DockQ scores of the top-1 and top-5 candidates were assessed.

As shown in Figure \ref{fig:fig_capri}, LambdaDockScore outperforms the baseline DFMDock model at top-1 and top-5 ranking of acceptable and medium quality poses, though it underperforms or matches baseline performance in ranking high quality poses. LambdaDockScore also outperforms state-of-the-art EuDockScore at top-1 and top-5 ranking of all three pose quality thresholds.

To analyze how performance varies with the number of top-ranked poses retained, we plot the success rate against $k$ for $k$ up to 100 on the CAPRI score set (Figure \ref{fig:fig_topk_capri}). LambdaDockScore exceeds EuDockScore at every $k$ for acceptable and medium quality poses. It also matches or exceeds baseline DFMDock at nearly every $k$ for these two quality thresholds, with particularly large gains at small $k$. This indicates that LambdaDockScore's performance is not tied to a particular $k$ cutoff. The performance gap at small $k$ suggests that the fine-tuning approach is particularly helpful for improving performance of the top few poses, as suggested by the LambdaLoss's rank-weighted difference term.

\subsection{LambdaDockScore improves docking performance on antibody-antigen complexes and protein complexes with extreme binding interface sizes}

\begin{figure}[h]
  \centering
  \includegraphics[width=0.85\textwidth]{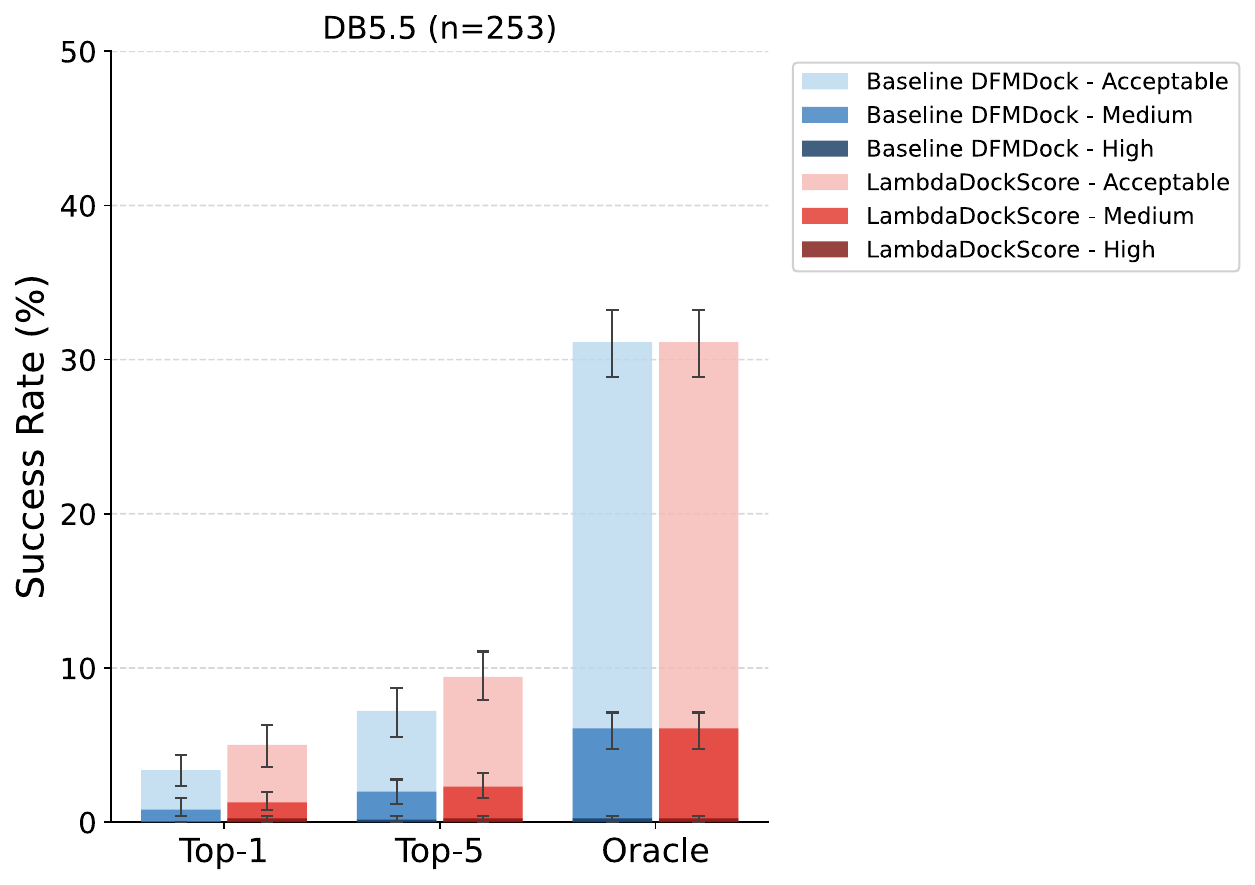} 
  \caption{Benchmarking results on 253 protein-protein complexes from Docking Benchmark 5.5. 120 decoys were generated per protein-protein complex using baseline DFMDock. The top-1 and top-5 ranked decoys from each model were analyzed. DFMDock with LambdaDockScore as its ranking model outperforms baseline DFMDock on both top-1 and top-5 ranking accuracy for acceptable and medium quality poses. Bar heights show the mean performance from 10,000 bootstrap samples, and error bars show the 95\% confidence interval. Exact success rates and 95\% confidence intervals are provided in Supplementary Table \ref{tab:db55_bar_heights}.}
  \label{fig:fig_db55}
\end{figure}

In Figure \ref{fig:fig_db55}, we compare the pose ranking performance of LambdaDockScore and baseline DFMDock energy prediction head for docking poses sampled with the baseline DFMDock model on the DB5.5 dataset. A total of 120 decoy poses per protein-protein complex in the dataset were generated, and then either the baseline energy prediction head or LambdaDockScore was used to rank the poses. The top-1, top-5, and oracle success rates are calculated, where the oracle is the maximum possible success rate given a perfect ranking of the sampled poses.

LambdaDockScore outperforms baseline DFMDock energy prediction head's top-1 and top-5 performance for both acceptable and medium quality poses. In Figure \ref{fig:fig_topk_db55}, we also look at both models' performance on DB5.5 as the number of top-$k$ poses is increased, up to $k=100$. Similar to the CAPRI score set test case, we observe that LambdaDockScore matches or exceeds baseline performance on most $k$ values considered,  with particularly large performance differences at small $k$ for acceptable quality poses. Therefore, the performance of DFMDock can be improved by combining its sampling process with the fine-tuned LambdaDockScore for ranking.

\begin{figure}[h]
  \centering
  \includegraphics[width=0.85\textwidth]{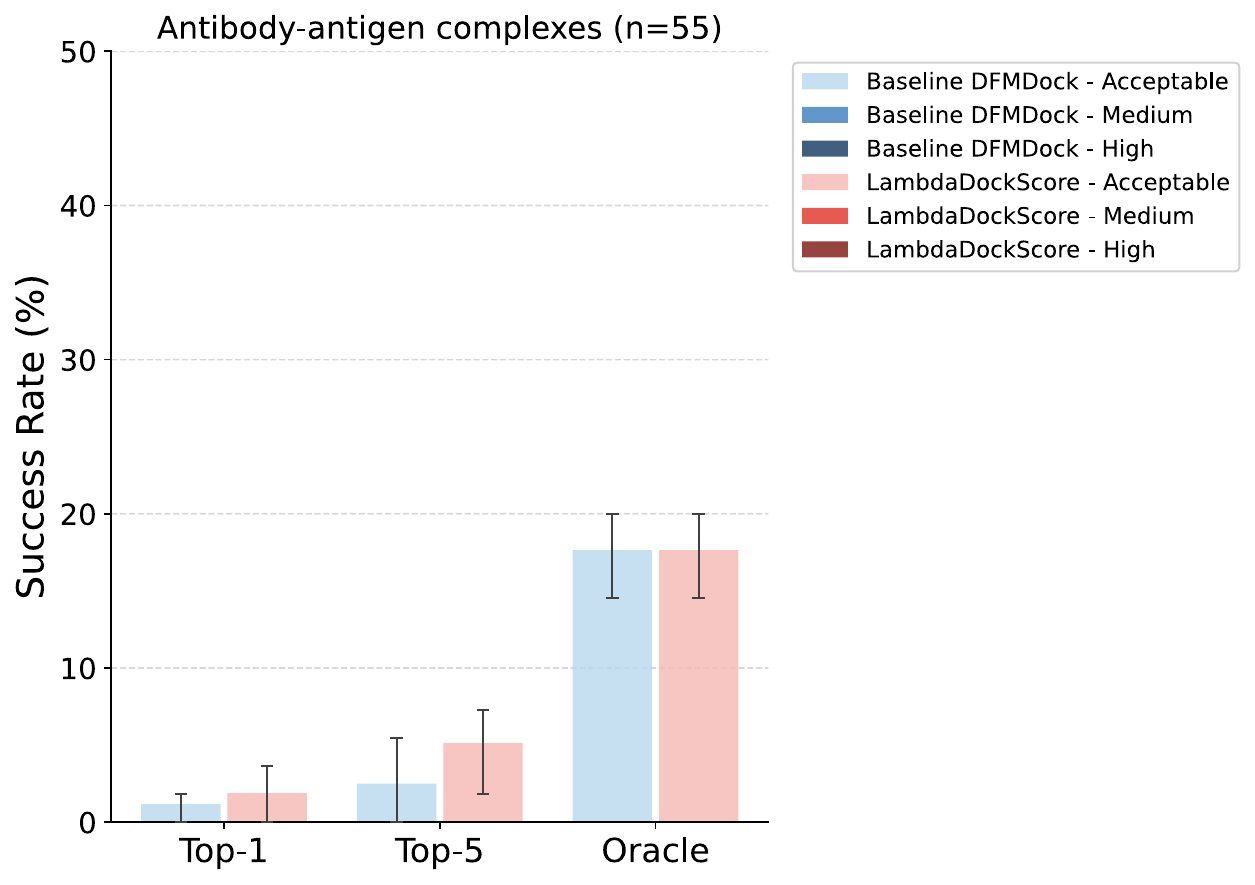} 
  \caption{Benchmarking results on antibody-antigen complexes from Docking Benchmark 5.5. Results are for 120 decoys generated and ranked per protein-protein complex. DFMDock with LambdaDockScore as its ranking model (red) outperforms baseline DFMDock (blue). Exact success rates and 95\% confidence intervals are provided in Supplementary Table \ref{tab:abag_bar_heights}.}
  \label{fig:fig_abantigen}
\end{figure}

Next, we analyze LambdaDockScore's performance on specific subtypes of protein-protein complexes within the DB5.5 benchmark set. Figure \ref{fig:fig_abantigen} shows that using LambdaDockScore improves baseline DFMDock's performance on antibody-antigen complexes in the DB5.5 benchmark dataset. Success rate improves from 1.2\% to 1.9\% for top-1 predictions and from 2.5\% to 5.1\% for top-5 predictions. Given the challenging nature of modeling antibody-antigen complexes, it is unsurprising that the absolute success rate percentage is low, as the oracle success rate (the maximum possible success rate) is only 17.6\%. Since the success rate is a function of both the sampling performance (with baseline DFMDock) and ranking performance (with LambdaDockScore), using a better baseline sampling method than DFMDock may further improve the overall performance on these difficult protein-protein complexes.

Finally, Supplementary Figure \ref{fig:fig_interface} shows LambdaDockScore's performance on DB5.5 segmented by the size of the binding interface as measured by the change in accessible surface area upon binding ($\Delta$ASA). We show that the performance improvement for ranking acceptable and medium quality poses is the greatest for protein-protein complexes in the bottom 40\% (Supplementary Figure \ref{fig:fig_interface}a-b) and in the top 20\% of interface sizes (Supplementary Figure \ref{fig:fig_interface}e), while the performance drops in the middle 40\% of protein-protein complex interface sizes (Supplementary Figure \ref{fig:fig_interface}c-d). Thus, fine-tuning with LambdaLoss allows the pose ranking model to better rank protein-protein complexes with extreme-sized (very small or very large) binding interfaces, with the trade-off of decreased performance on average-sized binding interfaces.

We note, however, that the middle interface-size bins are also the least rankable subsets of the benchmark. As shown in Supplementary Table \ref{tab:interface_rankability}, 36 of 51 complexes in the 40--60th percentile bin ($70.6\%$) and 34 of 50 complexes in the 60--80th percentile bin ($68\%$) contain no acceptable pose among the sampled decoys. The lower number of rankable complexes in the middle interface-size bins means that LambdaDockScore's performance in these bins should be interpreted with greater caution. Further analysis on larger and more rankable subsets is required.

\section{Conclusion}

In this study, we present LambdaDockScore, a novel method for improving protein-protein complex scoring functions using the LambdaLoss ranking loss. We show that by augmenting the original training data with carefully generated decoys and applying the LambdaLoss when fine-tuning the original ranking model on these decoys, we can not only improve the pose ranking accuracy beyond baseline, but also outperform EuDockScore, a state-of-the-art protein-protein complex scoring function. We also demonstrate that this data-augmentation-based LambdaLoss fine-tuning strategy specifically improves pose ranking performance on complexes with very large or very small binding interfaces, as well as antibody-antigen complexes. Improving performance on these challenging subcategories underscores the potential of LambdaDockScore. This LambdaLoss-based data augmentation and fine-tuning strategy could be used to improve numerous pose ranking problems and techniques in biomolecular modeling, such as the confidence modules of Boltz and AlphaFold. This would allow for improvements in not just protein-protein interactions, but also the grand challenge of modeling all biomolecular interactions.
\section*{Acknowledgements}

This work was supported by National Science Foundation REU Grant 2244288 and National Institutes of Health grant R35-GM141881. Computational resources were provided by the Advanced Research Computing at Hopkins (ARCH).

\section*{Code availability}

The code for this project is publicly available at \url{https://github.com/Graylab/LambdaDockScore}.

\section{References}

{
\small

[1] Abramson, J., Adler, J., Dunger, J., Evans, R., Green, T., Pritzel, A., Ronneberger, O., Willmore, L., Ballard, A. J., Bambrick, J., Bodenstein, S. W., Evans, D. A., Hung, C.-C., O’Neill, M., Reiman, D., Tunyasuvunakool, K., Wu, Z., Žemgulytė, A., Arvaniti, E., … Jumper, J. M. (2024). Accurate structure prediction of biomolecular interactions with AlphaFold 3. \textit{Nature, 630}(8016), 493–500. https://doi.org/10.1038/s41586-024-07487-w

[2] Basu, S., \& Wallner, B. (2016). DockQ: A Quality Measure for Protein-Protein Docking Models. \textit{PLOS One, 11}(8), e0161879. https://doi.org/10.1371/journal.pone.0161879

[3] Burges, C. J. C., Ragno, R., \& Le, Q. V. (2007). Learning to Rank with Nonsmooth Cost Functions. \textit{Advances in Neural Information Processing Systems 19}, 193–200. https://doi.org/10.7551/mitpress/7503.003.0029

[4] Chu, L.-S., Sarma, S., Xu, D., \& Gray, J. J. (2026). Unified Sampling and Ranking for Protein Docking With DFMDock. \textit{Proteins: Structure, Function, and Bioinformatics}, 1–10. https://doi.org/10.1002/prot.70160

[5] Hitawala, F. N., \& Gray, J. J. (2025). What does AlphaFold3 learn about antibody and nanobody docking, and what remains unsolved? \textit{mAbs, 17}(1), 2545601. https://doi.org/10.1080/19420862.2025.2545601

[6] Ketata, M. A., Laue, C., Mammadov, R., Stärk, H., Wu, M., Corso, G., Marquet, C., Barzilay, R., \& Jaakkola, T. S. (2023, April 8). DiffDock-PP: Rigid Protein-Protein Docking with Diffusion Models. \textit{The Eleventh International Conference on Learning Representations (ICLR)}. https://doi.org/10.48550/arXiv.2304.03889

[7] Lensink, M. F., Méndez, R., \& Wodak, S. J. (2007). Docking and scoring protein complexes: CAPRI 3rd Edition. \textit{Proteins: Structure, Function, and Bioinformatics, 69}(4), 704–718. https://doi.org/10.1002/prot.21804

[8] Lensink, M. F., \& Wodak, S. J. (2014). Score\_set: A CAPRI benchmark for scoring protein complexes: A Benchmark for Scoring Protein Complexes. \textit{Proteins: Structure, Function, and Bioinformatics, 82}(11), 3163–3169. https://doi.org/10.1002/prot.24678

[9] Liu, T.-Y. (2009). Learning to Rank for Information Retrieval. \textit{Foundations and Trends® in Information Retrieval, 3}(3), 225–331. https://doi.org/10.1561/1500000016

[10] McFee, M., Kim, J., \& Kim, P. M. (2024). EuDockScore: Euclidean graph neural networks for scoring protein–protein interfaces. \textit{Bioinformatics, 40}(11). https://doi.org/10.1093/bioinformatics/btae636

[11] Morehead, A., Chen, C., Sedova, A., \& Cheng, J. (2023). DIPS-Plus: The enhanced database of interacting protein structures for interface prediction. \textit{Scientific Data, 10}(1), 509. https://doi.org/10.1038/s41597-023-02409-3

[12] Park, M., Nett, R., Petersen, B., \& Sivasubramanian, A. (2026). Benchmarking antibody-antigen co-folding on human monomeric antigens. \textit{bioRxiv}. https://doi.org/10.64898/2026.08.20.745880

[13] Passaro, S., Corso, G., Wohlwend, J., Reveiz, M., Thaler, S., Somnath, V. R., Getz, N., Portnoi, T., Roy, J., Stark, H., Kwabi-Addo, D., Beaini, D., Jaakkola, T., \& Barzilay, R. (2025). Boltz-2: Towards Accurate and Efficient Binding Affinity Prediction. \textit{bioRxiv}. https://doi.org/10.1101/2025.06.14.659707

[14] Shirali, A., Stebliankin, V., Karki, U., Shi, J., Chapagain, P., \& Narasimhan, G. (2025). A comprehensive survey of scoring functions for protein docking models. \textit{BMC Bioinformatics, 26}(1), 25. https://doi.org/10.1186/s12859-024-05991-4

[15] Stebliankin, V., Shirali, A., Baral, P., Shi, J., Chapagain, P., Mathee, K., \& Narasimhan, G. (2023). Evaluating protein binding interfaces with transformer networks. \textit{Nature Machine Intelligence, 5}(9), 1042–1053. https://doi.org/10.1038/s42256-023-00715-4

[16] Sverrisson, F., Feydy, J., Correia, B. E., \& Bronstein, M. M. (2021). Fast end-to-end learning on protein surfaces. \textit{2021 IEEE/CVF Conference on Computer Vision and Pattern Recognition (CVPR)}, 15267–15276. https://doi.org/10.1109/CVPR46437.2021.01502

[17] Townshend, R. J. L., Bedi, R., Suriana, P. A., \& Dror, R. O. (2019, December 26). End-to-End Learning on 3D Protein Structure for Interface Prediction. \textit{33rd Conference on Neural Information Processing Systems (NeurIPS 2019)}. https://doi.org/10.48550/arXiv.1807.01297

[18] Vreven, T., Moal, I. H., Vangone, A., Pierce, B. G., Kastritis, P. L., Torchala, M., Chaleil, R., Jiménez-García, B., Bates, P. A., Fernandez-Recio, J., Bonvin, A. M. J. J., \& Weng, Z. (2015). Updates to the Integrated Protein–Protein Interaction Benchmarks: Docking Benchmark Version 5 and Affinity Benchmark Version 2. \textit{Journal of Molecular Biology, 427}(19), 3031–3041. https://doi.org/10.1016/j.jmb.2015.07.016

[19] Wang, X., Li, C., Golbandi, N., Bendersky, M., \& Najork, M. (2018). The LambdaLoss Framework for Ranking Metric Optimization. \textit{Proceedings of the 27th ACM International Conference on Information and Knowledge Management}, 1313–1322. https://doi.org/10.1145/3269206.3271784

[20] Yin, R., \& Pierce, B. G. (2024). Evaluation of AlphaFold antibody–antigen modeling with implications for improving predictive accuracy. \textit{Protein Science,} 33(1), e4865. https://doi.org/10.1002/pro.4865

}


\appendix

\renewcommand{\thefigure}{S\arabic{figure}}
\setcounter{figure}{0} 

\renewcommand{\thetable}{S\arabic{table}}
\setcounter{table}{0} 

\section{Supplementary Material}

\subsection{Details on the generation process and DockQ distribution of the fine-tuning dataset}

\begin{table}[h]
  \centering
  \begin{tabular}{lccc}
    \toprule
    Decoy set & Translation std. dev. (Å) & Rotation std dev. (°) & \# poses \\
    \midrule
    Perturbed GT — small  & 0.1  & 1.0   & 20  \\
    Perturbed GT — medium & 1.0  & 7.0   & 100 \\
    Perturbed GT — large  & 2.5  & 20.0  & 100 \\
    DFMDock samples   & N/A  & N/A   & 50  \\
    \midrule
    Total (per GT complex)   &      &       & 270 \\
    \bottomrule
  \end{tabular}

  \vspace{1ex}
  \vspace{1ex}
  \caption{Decoy pose sampling details. GT stands for ground truth. Standard deviations are for Normal distributions centered at 0, which were used to sample random translations and rotations for the perturbed GT poses.}
  \label{tab:decoy-distribution}
\end{table}

\begin{figure}[h]
  \centering
  \includegraphics[width=0.8\textwidth]{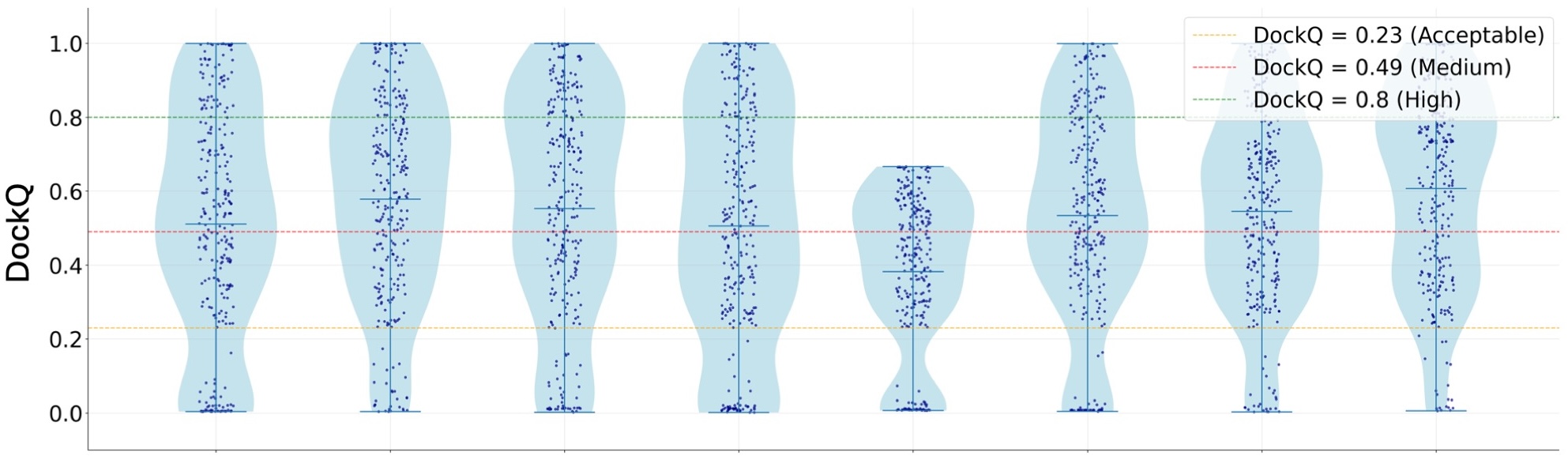} 
  \caption{Distribution of poses for eight randomly sampled protein-protein complexes from the DIPS-hetero fine-tuning dataset. The majority of complexes in the fine-tuning dataset have a distribution of poses across the full DockQ [0,1] interval, though some complexes lack sufficient numbers of high DockQ poses.} 
  \label{fig:fig_datadist}
\end{figure}
\newpage
\subsection{Success rate as a function of $k$, the number of top-ranked poses considered}

\begin{figure}[h]
  \centering
  \includegraphics[width=\textwidth]{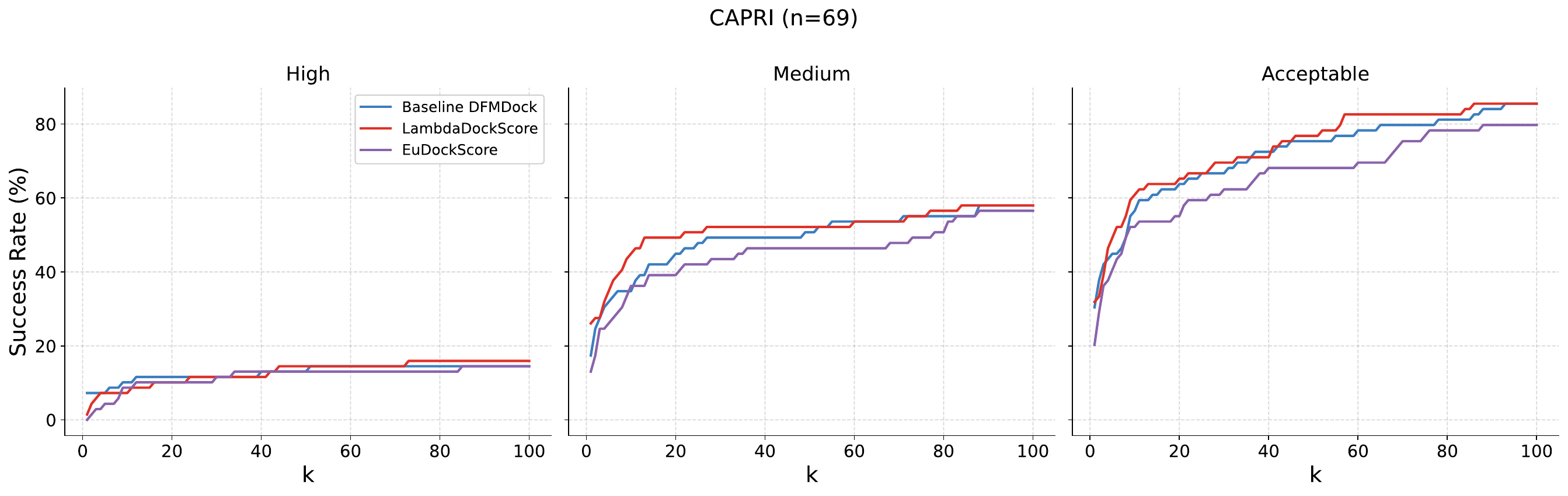}
  \caption{Top-$k$ success rate on the CAPRI score set (69 complexes) for LambdaDockScore (red), baseline DFMDock (blue), and EuDockScore (purple), shown for high (left), medium (center), and acceptable (right) quality thresholds. Each curve is the percentage of complexes with at least one pose above the quality threshold among the $k$ lowest-energy poses.}
  \label{fig:fig_topk_capri}
\end{figure}

\begin{figure}[h]
  \centering
  \includegraphics[width=\textwidth]{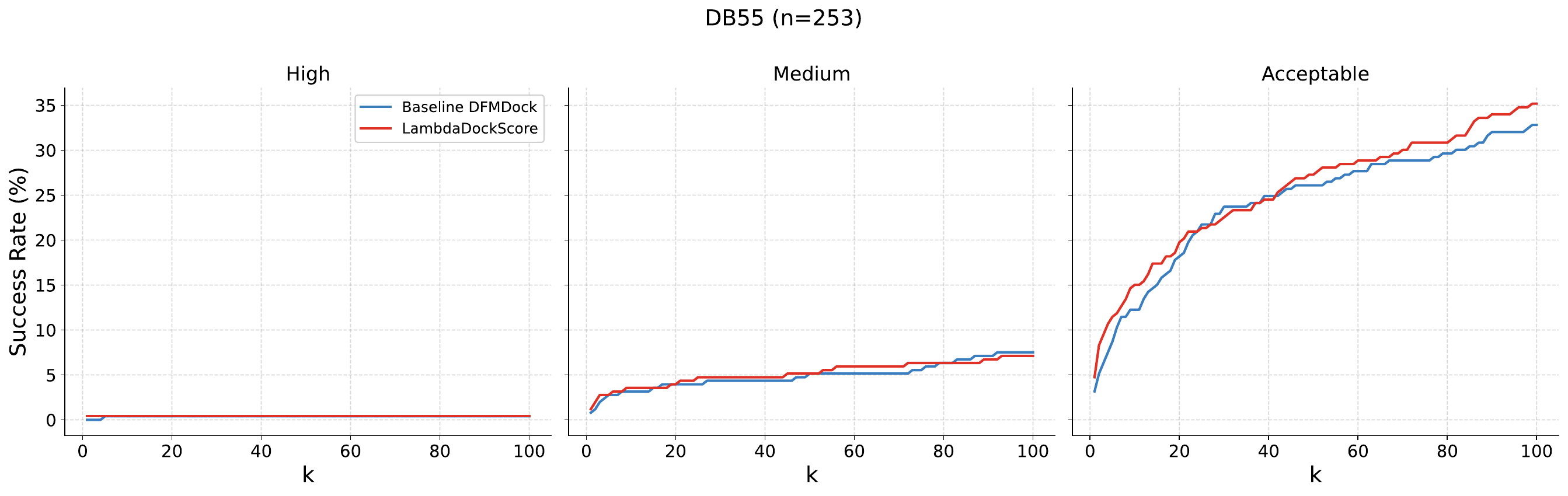}
  \caption{Top-$k$ success rate on DB5.5 (253 complexes) for LambdaDockScore (red) and baseline DFMDock (blue), shown for high (left), medium (center), and acceptable (right) quality thresholds. Poses were sampled with baseline DFMDock and ranked by each model. Each curve is the percentage of complexes with at least one pose above the quality threshold among the $k$ lowest-energy poses.}
  \label{fig:fig_topk_db55}
\end{figure}
\newpage

\subsection{LambdaDockScore performance on DB5.5, segmented by the size of the binding interface}
\begin{figure}[h]
  \centering

\includegraphics[width=0.9\textwidth]{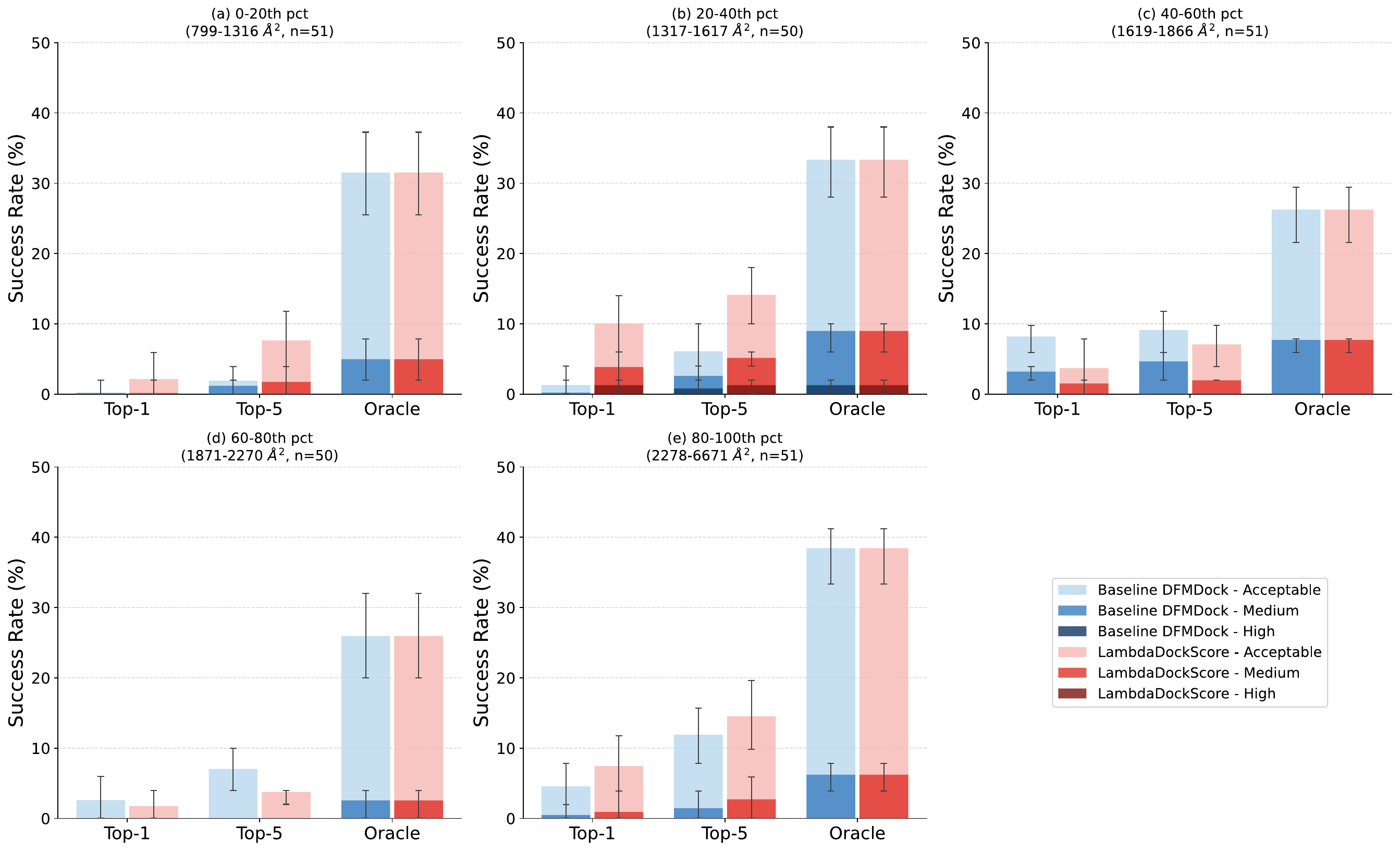} 
  \caption{Benchmarking results on Docking Benchmark 5.5, segmented by the change in accessible surface area ($\Delta$ASA). This is a measure of how the model performs based on the size of the binding interface. Protein-protein complexes were ordered by $\Delta$ASA and split into five evenly sized bins based on their $\Delta$ASA percentile. Panels \textbf{(a)-(e)} show the performance of baseline DFMDock's energy prediction head (blue) compared to LambdaDockScore (red) for each bin, in order of increasing interface size. Exact success rates and 95\% confidence intervals are provided in Supplementary Table \ref{tab:asa_bar_heights}.}
  \label{fig:fig_interface}
\end{figure}
\subsection{Rankability of DB5.5 complexes across $\Delta$ASA interface-size bins}
\begin{table}[h]
\centering
\begin{tabular}{lccc}
\toprule
$\Delta$ASA bin & Targets & Targets with zero acceptable poses & Percent with zero acceptable poses \\
\midrule
0--20th   & 51 & 32 & 62.7\% \\
20--40th  & 50 & 30 & 60\% \\
40--60th & 51 & 36 & 70.6\% \\
60--80th  & 50 & 34 & 68\% \\
80--100th & 51 & 30 & 58.8\% \\
\bottomrule
\end{tabular}
\caption{The number of target complexes in each $\Delta$ASA percentile bin, the number containing no acceptable pose among the sampled decoys, and the percent of target complexes with no acceptable pose.}
\label{tab:interface_rankability}
\end{table}

\newpage
\subsection{Tabulated Success Rates and Confidence Intervals}
\label{sec:supp_bar_heights}

This section provides the exact bar heights (success rates in \%) and bootstrap 95\% confidence intervals corresponding to the bar plots in the main text and supplementary figures. In all figures, bars are overlaid (Acceptable encompasses Medium, which encompasses High).

\begin{table}[H]
  \centering
  \caption{Success rates for the CAPRI Score Set (n=69), corresponding to Figure \ref{fig:fig_capri}.}
  \label{tab:capri_bar_heights}
  \begin{tabular}{llccc}
    \toprule
    Evaluation Metric & Model & Acceptable (\%) & Medium (\%) & High (\%) \\
    \midrule
    Top-1 & Baseline DFMDock & 30.4 & 17.4 & 7.2 \\
    Top-1 & LambdaDockScore & 31.9 & 26.1 & 1.4 \\
    Top-1 & EuDockScore & 20.3 & 13.0 & 0.0 \\
    Top-5 & Baseline DFMDock & 44.9 & 31.9 & 7.2 \\
    Top-5 & LambdaDockScore & 49.3 & 34.8 & 7.2 \\
    Top-5 & EuDockScore & 40.6 & 26.1 & 4.3 \\
    \bottomrule
  \end{tabular}
\end{table}

\begin{table}[H]
  \centering
  \caption{Success rates and 95\% confidence intervals for DB5.5 (n=253), corresponding to Figure \ref{fig:fig_db55}. Success rate values represent the mean performance from 10,000 pose-bootstrap samples.}
  \label{tab:db55_bar_heights}
  \resizebox{\textwidth}{!}{
  \begin{tabular}{llccc}
    \toprule
    Evaluation Metric & Model & Acceptable (\%) [95\% CI] & Medium (\%) [95\% CI] & High (\%) [95\% CI] \\
    \midrule
    Top-1 & Baseline DFMDock & 3.4 [2.4, 4.3] & 0.8 [0.4, 1.6] & 0.0 [0.0, 0.0] \\
    Top-1 & LambdaDockScore & 5.0 [3.6, 6.3] & 1.3 [0.8, 2.0] & 0.3 [0.0, 0.4] \\
    Top-5 & Baseline DFMDock & 7.2 [5.5, 8.7] & 2.0 [1.2, 2.8] & 0.2 [0.0, 0.4] \\
    Top-5 & LambdaDockScore & 9.4 [7.9, 11.1] & 2.3 [1.6, 3.2] & 0.3 [0.0, 0.4] \\
    Oracle & Baseline DFMDock & 31.1 [28.9, 33.2] & 6.1 [4.7, 7.1] & 0.3 [0.0, 0.4] \\
    Oracle & LambdaDockScore & 31.1 [28.9, 33.2] & 6.1 [4.7, 7.1] & 0.3 [0.0, 0.4] \\
    \bottomrule
  \end{tabular}
  }
\end{table}

\begin{table}[H]
  \centering
  \caption{Success rates and 95\% confidence intervals for DB5.5 Antibody-Antigen Complexes (n=55), corresponding to Figure \ref{fig:fig_abantigen}. Success rate values represent the mean performance from 10,000 pose-bootstrap samples.}
  \label{tab:abag_bar_heights}
  \resizebox{\textwidth}{!}{
  \begin{tabular}{llccc}
    \toprule
    Evaluation Metric & Model & Acceptable (\%) [95\% CI] & Medium (\%) [95\% CI] & High (\%) [95\% CI] \\
    \midrule
    Top-1 & Baseline DFMDock & 1.2 [0.0, 1.8] & 0.0 [0.0, 0.0] & 0.0 [0.0, 0.0] \\
    Top-1 & LambdaDockScore & 1.9 [0.0, 3.6] & 0.0 [0.0, 0.0] & 0.0 [0.0, 0.0] \\
    Top-5 & Baseline DFMDock & 2.5 [0.0, 5.5] & 0.0 [0.0, 0.0] & 0.0 [0.0, 0.0] \\
    Top-5 & LambdaDockScore & 5.1 [1.8, 7.3] & 0.0 [0.0, 0.0] & 0.0 [0.0, 0.0] \\
    Oracle & Baseline DFMDock & 17.6 [14.5, 20.0] & 0.0 [0.0, 0.0] & 0.0 [0.0, 0.0] \\
    Oracle & LambdaDockScore & 17.6 [14.5, 20.0] & 0.0 [0.0, 0.0] & 0.0 [0.0, 0.0] \\
    \bottomrule
  \end{tabular}
  }
\end{table}

\begin{table}[H]
  \centering
  \caption{Success rates and 95\% confidence intervals for DB5.5 split into 5 $\Delta$ASA bins, corresponding to Supplementary Figure \ref{fig:fig_interface}. Success rate values represent the mean performance from 10,000 pose-bootstrap samples across 5 interface area bins.}
  \label{tab:asa_bar_heights}
  \resizebox{\textwidth}{!}{
  \begin{tabular}{lllccc}
    \toprule
    $\Delta$ASA Bin & Metric & Model & Acceptable (\%) [95\% CI] & Medium (\%) [95\% CI] & High (\%) [95\% CI] \\
    \midrule
    \multirow{6}{*}{\shortstack[l]{\textbf{0--20th pct} \\ (799-1316 \AA$^2$ \\ n=51)}} 
    & Top-1 & Baseline DFMDock & 0.2 [0.0, 2.0] & 0.2 [0.0, 2.0] & 0.0 [0.0, 0.0] \\
    & Top-1 & LambdaDockScore & 2.1 [0.0, 5.9] & 0.2 [0.0, 2.0] & 0.0 [0.0, 0.0] \\
    & Top-5 & Baseline DFMDock & 1.9 [0.0, 3.9] & 1.2 [0.0, 2.0] & 0.0 [0.0, 0.0] \\
    & Top-5 & LambdaDockScore & 7.6 [3.9, 11.8] & 1.7 [0.0, 3.9] & 0.0 [0.0, 0.0] \\
    & Oracle & Baseline DFMDock & 31.5 [25.5, 37.3] & 5.0 [2.0, 7.8] & 0.0 [0.0, 0.0] \\
    & Oracle & LambdaDockScore & 31.5 [25.5, 37.3] & 5.0 [2.0, 7.8] & 0.0 [0.0, 0.0] \\
    \midrule
    \multirow{6}{*}{\shortstack[l]{\textbf{20--40th pct} \\ (1317-1617 \AA$^2$ \\ n=50)}} 
    & Top-1 & Baseline DFMDock & 1.3 [0.0, 4.0] & 0.2 [0.0, 2.0] & 0.0 [0.0, 0.0] \\
    & Top-1 & LambdaDockScore & 10.0 [6.0, 14.0] & 3.8 [2.0, 6.0] & 1.3 [0.0, 2.0] \\
    & Top-5 & Baseline DFMDock & 6.1 [4.0, 10.0] & 2.6 [0.0, 4.0] & 0.8 [0.0, 2.0] \\
    & Top-5 & LambdaDockScore & 14.1 [10.0, 18.0] & 5.2 [4.0, 6.0] & 1.3 [0.0, 2.0] \\
    & Oracle & Baseline DFMDock & 33.3 [28.0, 38.0] & 9.0 [6.0, 10.0] & 1.3 [0.0, 2.0] \\
    & Oracle & LambdaDockScore & 33.3 [28.0, 38.0] & 9.0 [6.0, 10.0] & 1.3 [0.0, 2.0] \\
    \midrule
    \multirow{6}{*}{\shortstack[l]{\textbf{40--60th pct} \\ (1619-1866 \AA$^2$ \\ n=51)}} 
    & Top-1 & Baseline DFMDock & 8.2 [5.9, 9.8] & 3.2 [2.0, 3.9] & 0.0 [0.0, 0.0] \\
    & Top-1 & LambdaDockScore & 3.7 [0.0, 7.8] & 1.5 [0.0, 2.0] & 0.0 [0.0, 0.0] \\
    & Top-5 & Baseline DFMDock & 9.1 [5.9, 11.8] & 4.6 [2.0, 5.9] & 0.0 [0.0, 0.0] \\
    & Top-5 & LambdaDockScore & 7.1 [3.9, 9.8] & 2.0 [2.0, 2.0] & 0.0 [0.0, 0.0] \\
    & Oracle & Baseline DFMDock & 26.3 [21.6, 29.4] & 7.7 [5.9, 7.8] & 0.0 [0.0, 0.0] \\
    & Oracle & LambdaDockScore & 26.3 [21.6, 29.4] & 7.7 [5.9, 7.8] & 0.0 [0.0, 0.0] \\
    \midrule
    \multirow{6}{*}{\shortstack[l]{\textbf{60--80th pct} \\ (1871-2270 \AA$^2$ \\ n=50)}} 
    & Top-1 & Baseline DFMDock & 2.6 [0.0, 6.0] & 0.0 [0.0, 0.0] & 0.0 [0.0, 0.0] \\
    & Top-1 & LambdaDockScore & 1.8 [0.0, 4.0] & 0.0 [0.0, 0.0] & 0.0 [0.0, 0.0] \\
    & Top-5 & Baseline DFMDock & 7.0 [4.0, 10.0] & 0.0 [0.0, 0.0] & 0.0 [0.0, 0.0] \\
    & Top-5 & LambdaDockScore & 3.8 [2.0, 4.0] & 0.0 [0.0, 0.0] & 0.0 [0.0, 0.0] \\
    & Oracle & Baseline DFMDock & 25.9 [20.0, 32.0] & 2.5 [0.0, 4.0] & 0.0 [0.0, 0.0] \\
    & Oracle & LambdaDockScore & 25.9 [20.0, 32.0] & 2.5 [0.0, 4.0] & 0.0 [0.0, 0.0] \\
    \midrule
    \multirow{6}{*}{\shortstack[l]{\textbf{80--100th pct} \\ (2278-6671 \AA$^2$ \\ n=51)}} 
    & Top-1 & Baseline DFMDock & 4.5 [2.0, 7.8] & 0.5 [0.0, 2.0] & 0.0 [0.0, 0.0] \\
    & Top-1 & LambdaDockScore & 7.5 [3.9, 11.8] & 0.9 [0.0, 3.9] & 0.0 [0.0, 0.0] \\
    & Top-5 & Baseline DFMDock & 11.9 [7.8, 15.7] & 1.5 [0.0, 3.9] & 0.0 [0.0, 0.0] \\
    & Top-5 & LambdaDockScore & 14.5 [9.8, 19.6] & 2.7 [0.0, 5.9] & 0.0 [0.0, 0.0] \\
    & Oracle & Baseline DFMDock & 38.4 [33.3, 41.2] & 6.2 [3.9, 7.8] & 0.0 [0.0, 0.0] \\
    & Oracle & LambdaDockScore & 38.4 [33.3, 41.2] & 6.2 [3.9, 7.8] & 0.0 [0.0, 0.0] \\
    \bottomrule
  \end{tabular}
  }
\end{table}

\end{document}